\documentclass{article}
\pdfoutput=1

\usepackage{authblk}
\usepackage{blindtext}
\usepackage{graphicx}
\usepackage{algorithm}
\usepackage{amsmath}
\usepackage{amssymb}
\usepackage{xcolor}
\usepackage{ulem}
\usepackage{cprotect}
\usepackage{url}

\begin{document}


\title{On the emergence of Zipf's law in music}





\author[1]{Juan I. Perotti\thanks{juanpool@gmail.com}}
\author[1,2]{Orlando V. Billoni}
\affil[1]{{Instituto de F\'isica Enrique Gaviola (IFEG-CONICET)\\ Ciudad Universitaria, 5000 C\'ordoba, Argentina}}
\affil[2]{Facultad de Matem\'atica Astronom\'{\i}a,  F\'{\i}sica y Computaci\'on\\ Universidad Nacional de C\'ordoba\\ Ciudad Universitaria, 5000 C\'ordoba, Argentina}

\maketitle

\begin{abstract}
Zipf's law is found when the vocabulary of long written texts is ranked according to the frequency of word occurrences, establishing a power-law decay for the frequency {\it vs} rank relation.
This law is a robust statistical property observed even in ancient untranslated languages. Interestingly, this law seems to be also manifested in music records when several metrics---functioning as words in written texts---are used.
Even though music can be regarded as a language,
finding an accurate equivalent of the concept of words in music is difficult because it lacks a functional semantic. 
This raises the question of which is the appropriate choice of Zipfian units in music, which is extensive to other contexts where this law can emerge. 
In particular, this is still an open question in written texts, where several alternatives have been proposed as Zipfian units besides the canonical use of words.
Seeking to validate a natural election of Zipfian units in music, in this work we find that Zipf's law emerges when a combination of chords and notes are chosen as Zipfian units. 
Our results are grounded on a consistent analysis of the statistical properties of music and texts, complemented with theoretical considerations that combine different reference models, including a simple model inspired in the Lempel-Ziv compression algorithm that we have devised to explain the emergence of Zipf’s law as the consequence of languages evolving into more efficient forms of communication.
\end{abstract}




\section{
\label{sec:introduction}
INTRODUCTION
}

When the components or events of several strikingly different complex systems are ranked according to an importance score, the rank $r$ and the scores $f$ are found to be related by a power-law scaling $f \sim r^{-z}$, where the exponent $z$ is usually close to one~\cite{bak1996how}, although exceptions to $z\approx 1$ are also reported~\cite{baixeries2013evolution,bian2016scaling,dasilva2018lotka}.
This relation is associated to a power-law distribution of scores $P(f) \sim f^{-\gamma}$, whose complementary cumulative form $P(\geq f)\sim f^{-\alpha}$ with $\gamma=1+\alpha$ is commonly known as Pareto's distribution~\cite{Lu1991-solar-flares, Wheatland98AJ,Corral-PRL-2004, barabasi2005origin, blasius2009zipf,Perotti2013}.
Generally, the validity of the power-law scaling holds over a restricted range and is often influenced by system size effects~\cite{newmann-zipf}.
In the particular context of text analysis, the ranked events are word occurrences, the scores $f$ are the associated frequencies and the $f\sim r^{-z}$ scaling is called Zipf's law~\cite{zipf1949human}.

Zipf's law is a robust trait of many written languages~\cite{mehri2017}, including ancient untranslated languages~\cite{Smith2007Gzipf-meroitic}, codexs with an undeciphered writing system such as the Voynich's manuscript~\cite{Landini2001-voynich} and artificially created languages such as Esperanto~\cite{manaris2006-esperanto}.
On the other hand, the standard form of Zipf's law only partially fits the empirical data~\cite{montemurro2002new,zanette2005dynamics,Piantadosi2014review-zipf} and the details of the rank-frequency distribution $f(r)$ depends on the specific literary piece or corpus analyzed.
Hence, different corrections to the standard form of Zipf's law were studied~\cite{montemurro2002new,zanette2005dynamics,ausloos2014toward_arxiv,williams2015zipf,moreno-sanchez2016large-scale,altmann2016statistical,Chierichetti2017,gerlach2019testing}.
One of the simplest corrections is the Zipf-Mandelbrot (ZM) law $f \propto (r_z+r)^{-z}$, which was early proposed by Mandelbrot~\cite{mandelbrot1953}.
Exceptions to Zipf's law have been also reported.
For example, the rank-frequency distributions of Chinese texts is better fitted by a stretched exponential (SE) function than by a power-law~\cite{deng2014EPJBzipf-chinese}.
Seeking to explain such exceptions, a recent work~\cite{yan2018dependence} suggests that the shape of the rank-frequency distribution depends on how languages are encoded.
With a similar point of view, another work~\cite{williams2015zipf} extends the validity of Zipf's law by using phrases instead of words as the semantic units for statistical linguistic analysis.
All these observations open the question of which are the appropriate choice of lexical or Zipfian units for the statistical analysis of the different forms of language.

Music is generally accepted as another form of human language~\cite{zipf1949human}, and as such, it involves perceptually discrete elements displaying organization in the form of underlying patterns and regularities~\cite{serra2012measuring}.
At variance with spoken languages, however, the quest for a semantic meaning in music is more elusive.
A potential semantic meaning could be the capacity of music to communicate emotions.
For example, online experiments provide a quantitative analysis of the relationship between music and emotions~\cite{Egermann2009,Weth2015,schaefer2017music}, and neurology studies show that music can induce physiological indices of semantic processing~\cite{Koelsch2004}.
Nevertheless, results like these do not provide enough information for the identification of lexical or conceptual Zipfian units in the statistical analysis of musical pieces.
In spite of that, Zipf's law has been reported in music by using several metrics as the analogous of words in texts~\cite{zipf1949human,Manaris2005-zipf-music,zanette2006zipf}.
For instance, Manaris {\it et al.} found Zipfian like rank-frequency distributions of various measures in music such as pitch, duration, melodic intervals, and harmonic consonance~\cite{Manaris2005-zipf-music}. 
Similarly, Liu {\it et al.} found long-tailed distributions analyzing the distribution of pitch fluctuation in corpora of works of several classical composers~\cite{liu2013statistical-music}. 
Nevertheless, these authors were not focused on the identification of appropriate musical Zipfian units.
The only exceptions are the works by D.H. Zanette~\cite{zanette2006zipf} and M.B. del R\'io et al~\cite{beltrndelRo2008universality}.  
Zanette found a rank-frequency distribution following the Zipf-Mandelbrot law when notes are used as Zipfian units, while del R\'io found a beta distribution when the duration of notes is neglected.
Although these previous results are based on the analysis of a few relatively short musical compositions, they constitute interesting findings because notes can be considered the building blocks of music~\cite{beran2003statistics}.

From a general perspective, the identification of natural Zipfian units in music has two major meanings that are intrinsically interdependent. 
On one side, it establishes a parallelism between music and written languages.
On the other side, it supports the validity of Zipf's law as a phenotype of any written language.
Hence, the identification of appropriate Zipfian units in music conveys an interesting research problem.
Motivated by previous observations, in this work we present empirical evidence that supports the validity of Zipf's law in musical scores when a combination of chords and notes---here called {\em generalized chords}---are considered as Zipfian units.
Also, we show that Zipf's law fails for the case of notes on sufficiently large compositions, as previously found with similar metrics~\cite{beltrndelRo2008universality}.
Finally, considering that several mechanisms have been proposed to explain the emergence of Zipf's law~\cite{simon1955class, Li92randomtexts, Kanter95PRL-zipf-markov,cancho2003,Manin2009MandelbrotsMF,FerreriCancho2010random-text,thurner2015understanding}, in this work we introduce a simple mechanism---the so-called n-Gram Compression Algorithm (nGCA)---to motivate an explanation of why Zipf's law emerges for generalized chords but not for notes.
The presented results are endorsed by a comparative analysis combining written texts and reference models.

This paper is organized as follows.
Sec.~\ref{subsec:zipflaw_music} presents empirical evidence supporting the validity of Zipf's law in musical scores when generalized chords are used as Zipfian units.
It also includes corresponding evidence against notes.
Sec.~\ref{subsec:theory_compression} introduces an explanation of why Zipf's law emerges for generalized chords and not for notes.
The explanation is formalized in terms of the nGCA, which is based on information-theoretic arguments underlying the emergence of Zipf's law.
This section also presents a comparative analysis between musical and literary works.
Sec.~\ref{subsec:heaps} supplements the analysis of Zipf's law in music with the study of Heaps' law.
Finally, Sec.~\ref{sec:discussion} discusses the results and  Sec.~\ref{sec:conclusions} concludes.

\section{
\label{sec:results}
RESULTS
}

\subsection{
\label{subsec:zipflaw_music}
Analysis of Zipf's law in musical scores
}

\begin{figure}
\centering
\includegraphics{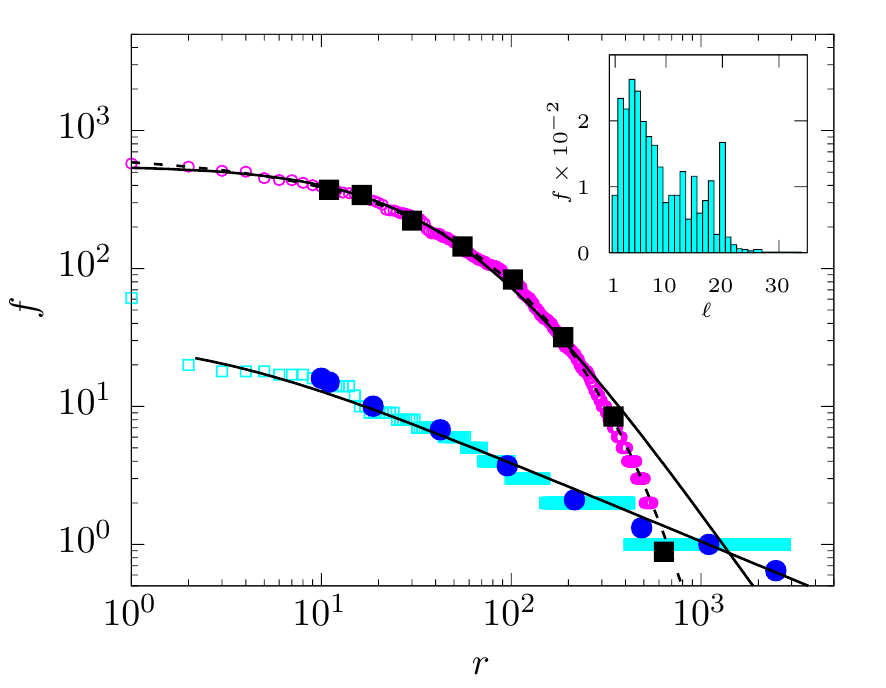}
\caption{
\label{fig:fig1} 
(Color online)
Frequency $f$ {\it vs} rank $r$ distributions obtained from a musical score of the II movement of Beethoven's Symphony No. 9:
in magenta open circles for notes and in cyan open squares for generalized chords (g-chords). 
The solid black squares and blue circles 
correspond to the logarithmic binning of previous curves, respectively.
The best-fitting Zipf-Mandelbrot (ZM) functions 
$f\sim (r+r_z)^{-z}$ 
are shown as solid lines, and result in
$z=1.98 \pm 0.03$ 
and
$r_z=56\pm 1$
for notes, and 
$z=0.572 \pm 0.003$ 
and
$r_z=2.5\pm 0.2$
for g-chords.
The best-fitting stretched exponential (SE) functions $f\sim \exp(-(r/r_0)^{\xi})$ are shown as dashed lines, resulting in
$r_0=26.5\pm 0.3$
and
$\xi=0.582\pm 0.003$ for notes, and 
$r_0=1\times 10^{-12}\pm 4\times 10^{-12}$
and
$\xi=0.07\pm 0.004$
for generalized chords.
Notice that in the case of g-chords, the SE fit results in unrealistically small values for $r_0$.
Inset: frequency distribution of g-chord lengths $\ell$, where $\ell=|g|$.
}
\end{figure}

\begin{figure}
\centering
\includegraphics{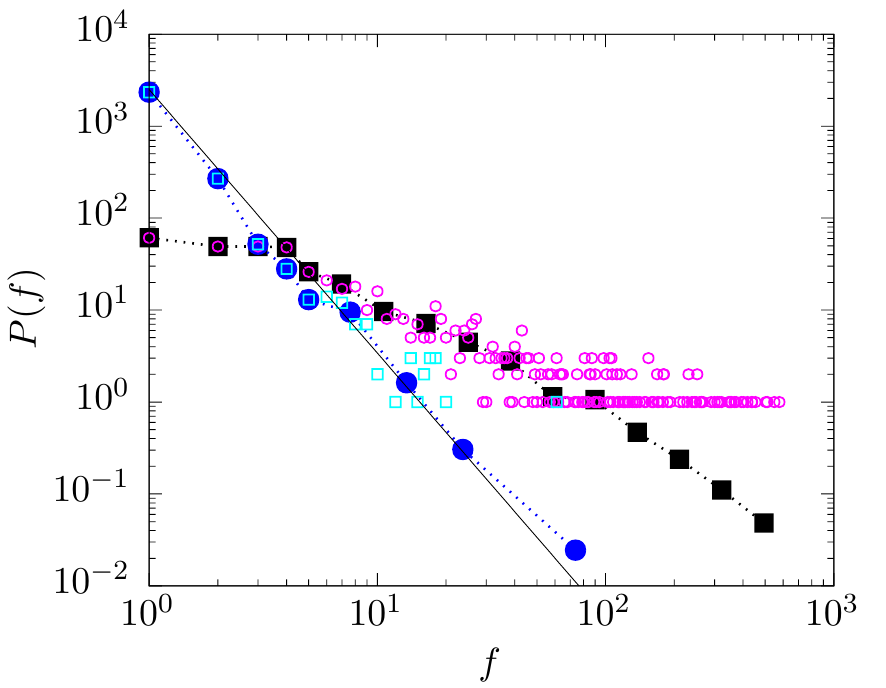}
\caption{
\label{fig:fig2} 
(Color online)
Distribution of frequencies $f$ for notes and g-chords.
Symbols and colors follow the convention of Fig.~\ref{fig:fig1}.
The solid line is a power-law guide to the eye of exponent $\gamma_{\mathrm{Hill}}=2.87$.
}
\end{figure}

From a general perspective, a musical piece consists of a complex sequence of combined sounds originated from different sources, such as musical instruments or singers.
The identification of sound attributes like duration, pitch, timbre, etc., enables the encoding of musical compositions into musical scores through the use of a specific notation.

Essentially, a musical score specifies when and how different notes and chords must be played by specific instruments as the interpretation of a musical piece develops in time.
A note is a pair $(d,p)$ specifying the duration $d$ and pitch $p$ of a sound.
A chord is a set of harmonically related notes played simultaneously by one instrument.
It is reasonable to think that chords represent a musical combination of notes analogous to the literary combination of characters into words.
Hence, chords seem a natural election of Zipfian units for the study of musical scores.
However, since the processing of chords is technically challenging, it is convenient to simplify musical scores into more manageable forms.
In doing so, we begin noticing that ordinary chords can be unfolded into their constituent notes, so a musical score can be reduced into a multi-set of time-stamped notes.
Moreover, the multi-set of time-stamped notes can be further condensed into a time respecting sequence $g_1,g_2,...$ of multi-sets $g$ of notes played simultaneously~(see~\ref{app:A} for details).
These multi-sets $g$ are similar to ordinary chords because they often contain several harmonically related notes. 
Still, the $g$s are different from plain chords because they may involve notes from different instruments.
We use the term {\em generalized chords}, or simply {\em g-chords}, to refer to these multi-sets $g$.
The sequences of g-chords conveniently simplify the original musical scores.
They summarize the information about single notes and chords, neglecting certain harmonic relations between the simultaneously played notes and the information about the instruments.

Given some arbitrary ordering over simultaneously played notes, a time respecting sequence of g-chords $g_1,g_2,...$ can be univocally mapped into an ordered sequence of notes $n_1,n_2,...$~(see~\ref{app:A}).
Taking advantage of the available string-format representation of notes, we sort them in alphabetical order.
It should be remarked, however, that any other ordering is equally artificial, but none of them affects the rank-frequency distribution.
Notwithstanding, some ordering for notes becomes useful in the ulterior analysis with the models and with Heaps' law.

Let us now explore Zipf's law in music by analyzing the frequency of occurrences of notes and g-chords in musical scores regardless of their time of appearance.
The musical scores here studied are available at~\cite{musescore} in \verb+.mxl+ format~\cite{mxlformat}, and can be pre-processed using \verb+Music21 Python+'s package~\cite{music21}.
We provide code examples in~\cite{perotti2019zipf_music_gitcode}.
For more details, please check~\ref{app:B}.
In Fig.~\ref{fig:fig1}, the frequency {\it vs} rank distributions for notes and g-chords are plotted for {\em Beethoven's Symphony No. 9, II movement}. 
The data for notes follows closely a SE function 
$f \sim e^{-(r/r_0)^{\xi}}$ 
as shown by the  best-fitting curve (dashed line). 
In contrast, the best-fitting ZM function $f \sim (r+r_z)^{-z}$ deviates at the tail (solid line).
For g-chords, no SE function provides a good fit of the distribution (not shown) since a non-realistic small value of $r_0$ is obtained.
Instead, the ZM function works much better, resulting in the best-fitting exponent $z \approx 0.6$ (solid curve). 
Please, note that we are not interested in the validation of Mandelbrot's model~\cite{Manin2009MandelbrotsMF}, for which a particular relation between $r_z$ and $z$ holds, but instead on the statistical validation of ZM distribution. 
We would like to mention, however, that the fitted $r_z$ yield reasonable values, following an exponential distribution with mean $\langle r_z\rangle \approx 3.9$ (not shown).
The inset depicts the distribution of g-chords sizes $\ell = |g|$ defined as the number of notes they contain.
Very large g-chords can be observed---up to $\ell \approx 30$---because, within a single symphony, there can be many instruments playing simultaneously.

We are aware that other Zipfian-like distributions could provide a more accurate description of the frequency statistics for g-chords than the ZM distribution.
Even more, non-Zipfian alternatives to SE distributions may work better for notes~\cite{beltrndelRo2008universality}.
However, we choose to work with ZM and SE functions because of several reasons: {\em i)} there is evidence supporting ZM law for music, {\em ii)} due to the limited statistics offered by the data, we must restrict ourselves to the distinction between Zipfian like and non-power-law statistics ignoring further refinements, {\em iii)} since the ZM and SE distributions have the same number of free parameters, the statistical significance of their fittings can be straightforwardly compared, and {\em iv)} the ZM is associated to a research line that is grounded in information-theoretic aspects that pave the way for the introduction of the nGCA. Further investigations using alternative distributions are left to be addressed in future works.

When a frequency {\it vs} rank curve follows a power-law scaling  $f\sim r^{-z}$, the distribution of frequencies also follows a power-law scaling $P(f) \sim f^{-\gamma}$ with 
$\gamma=1+1/z$~\cite{lu2010zipf}, a relation that holds at least asymptotically~\cite{corral2019distinct_arxiv}.
Fig.~\ref{fig:fig2} shows the frequency distributions associated to the frequency {\it vs} rank distributions of Fig.~\ref{fig:fig1}. 
In the case of g-chords, the distribution is well fitted by a power-law, while in the case of notes, the distribution seems to be short-tailed.
Also, for g-chords the power-law fit results in $\gamma=2.87$, while the relation $\gamma=1+1/z$ yields the value $\gamma=2.75$; both ways of inferring gamma are thus in agreement.

Next, we study the statistical robustness of the results shown in Figs.~\ref{fig:fig1}~and~\ref{fig:fig2}.
The statistical validation or rejection of the power-law scaling is more easily established for $P(f)$ than for $f(r)$~\cite{corral2019distinct_arxiv}.
In the present work, the statistical validation or rejection of a power-law scaling for the tail of $P(f)$ is determined using the test introduced by Voitalov {\it et al.}~\cite{voitalov2018scale}.
Voitalov's Test (VT) combines the concepts of regularly varying distributions and extreme value statistics, automatically taking into consideration the range of validity for the power-law scaling.
It rejects a power-law scaling if at least one of three different estimators $\gamma_{hill}$, $\gamma_{Moment}$ and $\gamma_{kernel}$ of $\gamma$ results $\geq 5$, and accepts it otherwise.
Of course, like any other approach, VT has limitations. 
For example, it fails for $z>1$, and thus there exists frequency distributions $P(f)$ that can be rejected even if a power-law scaling holds for $f(r)$.
Overall, we remark that VT is rather conservative, and therefore it constitutes a robust and up-to-date check for the existence of power-law scaling in frequency distributions.

\begin{figure}
\centering
\includegraphics{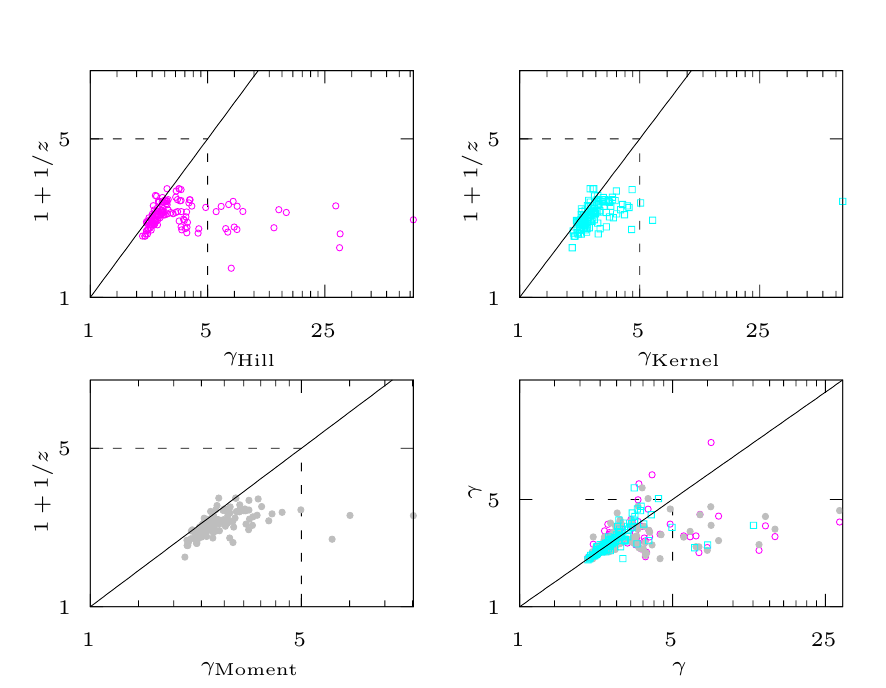}
\caption{
\label{fig:fig3} 
(Color online)
Voitalov's tests (VT) for the distribution of g-chord frequencies using {\em Hill}'s (top left), {\em Kernel} (top right) and {\em Moment} (bottom left) estimators of the $\gamma$ exponent, compared to the corresponding best-fitting ZM exponents $z$ in a corpus of 135 musical scores by Beethoven.
In the bottom right panel, pairs of estimators are compared against each other.
Open circles for {\em Hill} and {\em Moment}, solid circles for {\em Hill} and {\em Kernel}, and open squares for {\em Moment} and {\em Kernel}.
The solid lines denote the perfect ideal match.
According to the most conservative version of the VT, the power-law scaling for the distribution $P(f)\sim f^{\gamma}$ should be rejected if any of the estimators for $\gamma$ results in the condition $\gamma \geq 5$ or, in other words, When a point is lying outside of one of the dashed boxes.
An $\%83$ of musical scores pass VT, supporting the validity of the ZM law when g-chords are used as Zipfian units.
}
\end{figure}

\begin{figure}
\centering
\includegraphics{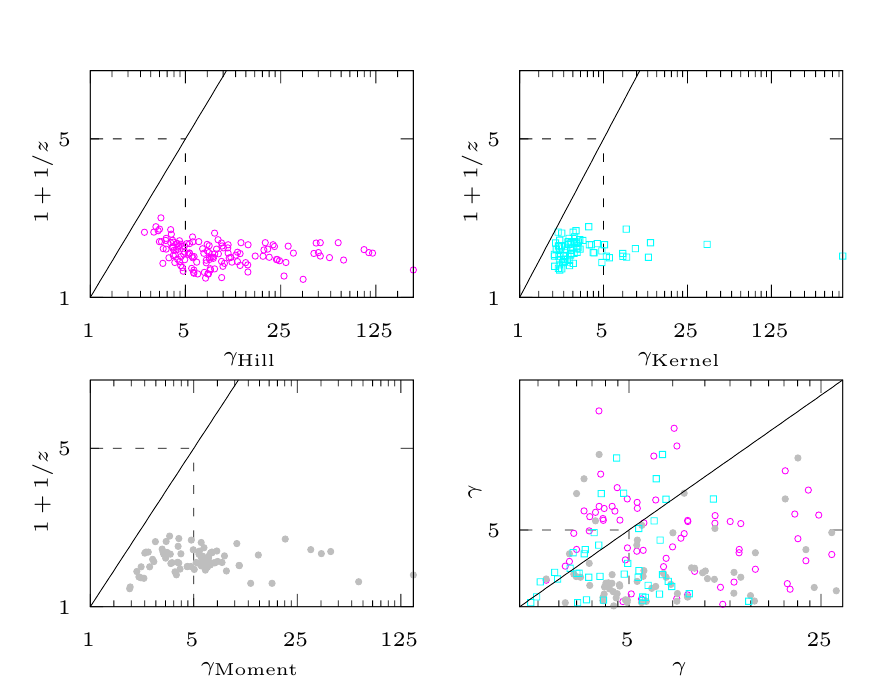}
\caption{
\label{fig:fig4} 
(Color online)
The VT for notes fails in nearly all the cases, since
only $\%3$ of the distributions pass the test.
Moreover, the expected relation $1+1/z \approx \gamma$ is clearly not fulfilled (top and bottom-left panels) and the correlation between the different $\gamma$ estimators is weak (bottom-right panel).
The statistical evidence indicates that the ZM law fails for notes.
}
\end{figure}

Based on the results of Figs.~\ref{fig:fig1}~and~\ref{fig:fig2}, VT is expected to work for g-chords---where typically $z\lesssim 1$ is observed, corresponding to $\gamma \gtrsim 2$---supporting their validity as proper Zipfian units for musical scores.
On the other hand, VT is expected to fail for notes, endorsing their rejection as Zipfian units.
Figs.~\ref{fig:fig3}~and~\ref{fig:fig4} summarize the results of VT applied to a corpus of $135$ Beethoven's musical scores.
The corpus includes complete symphonies, string quartets, sonatas for piano, etc.
All the music scores analyzed contain at least $5000$ notes.
Figs.~\ref{fig:fig3} (a), (b) and (c) show the results of VT for g-chords, where  $1+1/z$ is plotted against the different estimators for $\gamma$.
Here, $z$ is obtained from the best-fitting ZM function of the rank-frequency distribution.
Both measures should be equivalent for perfect power-laws, hence each plot (a), (b) and (c) is expected to follow a straight line. 
In all the cases, the curves approach a straight line in a range of relatively small values of $\gamma$.  
Also, most of the points are within the boxes defined by the dashed lines, meaning that a high percentage of the distributions pass VT. 
In Fig.~\ref{fig:fig3} (d), the different estimators are paired together, plotting them against each other.
All pairs are highly correlated when both estimators predict $\gamma < 5$.
For the case of g-chords, $\% 83$ of the distributions pass VT and the relation $\gamma\approx 1+1/z$ holds approximately in such cases.
The same analysis is repeated for notes in Fig.~\ref{fig:fig4} in panels (a-d), but with the majority of points lying outside of the boxes.
In this case, only $\% 3$ of the distributions pass VT, indicating the general failure of a power-law scaling for the case of notes.
Moreover, there is no good correlation between $1+1/z$
and the estimators of $\gamma$, nor between estimators themselves.
The previous results indicate that g-chords convey a good choice for Zipfian units and that the opposite holds for notes.

\subsection{
\label{subsec:theory_compression}
Zipf's law and communication efficiency
}

The idea that Zipf's law emerges as a consequence of optimization or adaptation processes where languages evolve into more efficient forms of communication is not new.
In fact, 
G.K. Zipf popularized it promoting the principle of least effort as an explanation for its emergence~\cite{zipf1949human}.
Later, B.B. Mandelbrot proposed a formalization of similar ideas borrowing formal concepts from information theory~\cite{mandelbrot1953,Manin2009MandelbrotsMF}.
This line of thinking initiated by Zipf and Mandelbrot is still active in the research of Zipf's law, with modern approaches combining theoretical and empirical efforts~\cite{montemurro2002new,cancho2003,baek2011zipf,seoane2018morphospace,dasilva2018granularity}. 
Alternative explanations also emerged, for instance, in terms of preferential growth processes~\cite{simon1955class,gabaix1999zipf,zanette2005dynamics,zanette2006zipf}.
It is important to remark, however, that it is unclear why such alternative processes could be at work, or if they necessarily invalidate Zipf-Mandelbrot ideas.
Perhaps all these alternative mechanisms were somehow evolutionarily selected for the ulterior purpose of language optimization.
Following these lines of thinking, let us analyze our results adhering to the idea that underlying the emergence of Zipf's law in languages, there are one or several mechanisms serving the optimization of encoding into more efficient forms of communication.
Specifically, we propose to model the compound effect of these mechanisms using an algorithm similar to the Lempel-Ziv information compressor~\cite{Gao2008, Montemurro2011-entropy}, which is an algorithm for the identification of repeating patterns within sequences of tokens that can be eventually encoded into shorter forms.

Let us introduce our proposed algorithm.
We call it the $n$-Gram Compression Algorithm or nGCA (see Algorithm~\ref{alg:nGCA}).
Later, we will focus on the optimization of the encoding of musical compositions into musical scores.
But, for the moment, we consider the general setting of a sequence $S$ of tokens.
The tokens could represent letters, words, notes, chords or any other predefined set of symbols.
Tuples of $n$ consecutive tokens in $S$ are identified as $n$-grams.
The nGCA works as follows.
First, $n$ is initialized to a sufficiently large value, so that no repeating $m$-gram exists in $S$ for $m > n$.
Then, the following procedure is iterated while $n>1$.
The set $P$ of all repeating $n$-grams of $S$ is computed.
If $P$ is empty, then $n$ is decreased by one and the next iteration follows.
Otherwise, a particular $n$-gram $q \in P$ is selected at random and is replaced by a compressed version $c_q$ which is defined through the concatenation of its tokens. 
Concatenation is accomplished with the help of some arbitrary and predefined special symbol.
For instance, if the special symbol is \verb+-+, $n=4$ and $q=$ \verb+(down,the,rabbit,hole)+, then $c_q=$  \verb+down-the-rabbit-hole+.
Then, following the order of appearance, each occurrence of $q$ in $S$ is replaced by $c_q$.
Since $q$ is a $n$-gram and $c_q$ is a 1-gram, the sequence $S$ shrinks and the alphabet expands after the replacement.
The algorithm stops when no repeating $n$-grams remain in $S$ for any $n$.
For details, please check a \verb+Python+ implementation of the algorithm~\cite{perotti2019zipf_music_gitcode} which, strictly speaking, produces no actual compression because, for practical and visualization purposes, our implementation skips the last step where the repeating $n$-grams, such as \verb+down-the-rabbit-hole+, are replaced by shorter tokens.
The identification of the compressible sub-sequences of tokens is sufficient for our purposes.

We now describe some interesting properties of the compressed sequences generated by the nGCA.
The unfolding of some arbitrary finite sequence $s_1,s_2,...$ can be approximately described by a random walk with memory over a co-occurrence directed multi-graph~\cite{mihalcea2011graph}, in which the tokens $s$ represent the nodes and the consecutive pairs of tokens $(s_i,s_{i+1})$ of the sequence represent the multiple directed links from node $s_i$ to node $s_{i+1}$.
The random walk approximation is essentially determined by the initial condition $s_1$ and a conditional probability $P(s_{i+1}|s_{i},...,s_{i+l-1})$ corresponding to a memory kernel of size $l$.
The size $l$ of the kernel bounds the goodness of the approximation.
For small values of $l$, the best approximation is generally bad. 
For a sufficiently large kernel, the sequence can be deterministically reproduced turning the approximation exact.
For a compressed sequence, the best approximating random walk model acquires special properties.
Namely, the approximation is maximally entropic for a kernel of size $l=1$ and deterministic for $l=2$.
Moreover, for $l=1$ the entropy rate at node $s$ equals $\ln k_s$, where $k_s$ is the out-degree of node $s$, and for $l=2$ the entropy rate is zero.
These properties arise because the algorithm fully compresses the repeating patterns within the sequence, turning the associated stochastic process into a deterministic Markovian chain.
Notice, however, that long-range memory effects may persist after compression since the original walk is generally non-ergodic, a fact related to the non-zero entropy rate of the original sequence~\cite{Gao2008}.
In other words, the nGCA only compresses the short-range memory effects caused by the repeating patterns found in the sequence, leaving unnoticed the long-range memory patterns~\cite{montemurro2002long-range,jafari2007detrended,boon2010complexity}, which cannot be captured by the nGCA due to the limited statistics offered by the data.

\begin{algorithm}[H]
  \begin{enumerate}
    \item[] {\bf input:} A sequence $S$.
    \item[] {\bf output:} The compressed version of $S$.
    \item[] Find the largest $n$ for which at least one $n$-gram in $S$ appears more than once.
    \item[] {\bf while} $n>1$
    \begin{enumerate}        
        \item[] Compute the set $P$ of the most frequent and repeating $n$-grams in $S$.
        \item[] {\bf if}  $P \neq \emptyset$ {\bf then}
        \begin{enumerate}
        \item[] Randomly select an $n$-gram $q$ from $P$.
        \item[] Let $c_q$ be the compressed version $c_q$ of $q$.
        \item[] Replace each occurrence of $q$ in $S$ by $c_q$.
        \end{enumerate}
        \item[] {\bf else}
        \begin{enumerate}
            \item[] $n\gets n-1$
        \end{enumerate}
    \end{enumerate}
    \item[] {\bf return} $S$
  \end{enumerate}
  \caption{The nGCA.}
  \label{alg:nGCA}
\end{algorithm}

Let us now illustrate how the nGCA works by applying it to a sequence of fragmented words. 
For the sake of definiteness, consider {\em Alice's adventures in wonderland}~\cite{gutenberg}---in short {\em Alice}---and replace each vowel of its words by the same vowel plus one space.
The procedure fragments every word in the text having an inner vowel.
For example, the beginning of a standardized version of {\em Alice}~\footnote{A version with all letters cast into lowercase form and with all punctuation signs removed.}
\begin{verbatim} 
down the rabbit hole alice was beginning to get very tired of 
sitting ...
\end{verbatim}
\noindent after fragmentation becomes,
\begin{verbatim} 
do wn the ra bbi t ho le a li ce wa s be gi nni ng to ge t ve ry 
ti re d o f si tti ng ...
\end{verbatim}
The application of the nGCA to the whole fragmented text of {\em Alice} results in
\begin{verbatim} 
do-wn the-ra-bbi-t-ho-le a-li-ce-wa-s-be-gi-nni-ng to-ge-t 
ve-ry-ti-re-d-o-f si-tti-ng ...
\end{verbatim}
As it can be seen, words such as \verb+down+ and \verb+sitting+ are fully recovered by the algorithm, since the corresponding patterns are repeated many times along the original text.
Additionally, new tokens not belonging to the original text, such as \verb+the-ra-bbi-t-ho-le+, are also found, but these tokens correspond to regularities most likely occurring in the particular literary work under consideration and not patterns proper of the English language.

\begin{figure}
\centering
\includegraphics{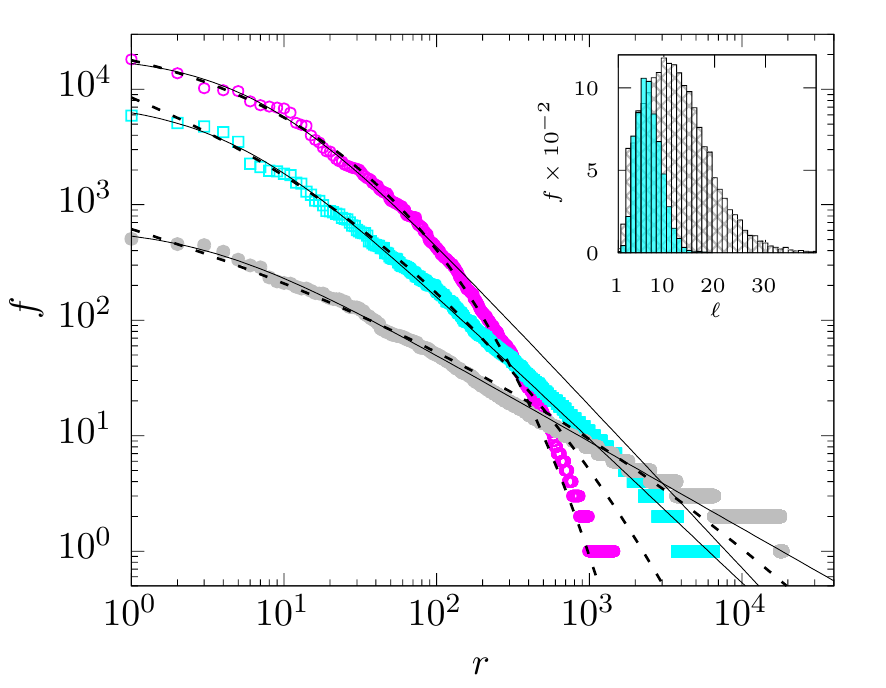}
\caption{
\label{fig:fig5} 
(Color online)
Frequency $f$ {\it vs} rank $r$ distributions for the literary work {\em Robinson Crusoe}.
Magenta open squares for the case of fragmented words (the analogous of notes) and cyan filled circles for the case of words (the analogous of g-chords).
Gray open squares for the compressed sequence of word-fragments.
The solid lines represent the best-fitting ZM functions with $z=1.41 \pm 0.01$ and $r_z=7.0 \pm 0.2$ for fragmented words, $z=1.248 \pm 0.005$ for ordinary words $r_z=4.49 \pm 0.05$, and $z=0.7533 \pm 0.0006$ and $r_z=3.35 \pm 0.02$ for the compressed sequence.
The dashed lines are the best-fitting SE functions with $\xi=0.363 \pm 0.004$ and $r_0=1.44 \pm 0.07$, $\xi=0.308 \pm 0.003$ and $r_0=0.53 \pm 0.03$, and $\xi=0.1 \pm 0.0007$ and $r_0=6\times 10^{-7} \pm 10^{-7}$, correspondingly.
Notice that the best-fitting ZM function for word-fragments is not satisfactory, similarly to the case of notes in Fig.~\ref{fig:fig1}.
Inset: length distributions of words and tokens of the compressed sequence in filled cyan and gray pattern, respectively.
}
\end{figure}

To study the statistical properties of the compressed sequences, let us now consider the literary work {\em Robinson Crusoe}~\cite{gutenberg}---a larger text than {\em Alice}---to comparatively analyze the emergence of Zipf's law using the previously introduced variation of the original text as reference models.
In Fig.~\ref{fig:fig5}, the frequency {\it vs} rank distributions is shown for the sequence of fragmented words, the original sequence, and the compressed sequence.
The fragmented case is well fitted by a SE function, whereas a ZM function performs poorly.
The opposite holds for the other two cases, where ZM function provides a better fit.
These results suggest that fragmented words behave as notes in musical scores, while words and compressed fragments of words behave analogously to g-chords.
The inset displays the distributions of words and compressed tokens lengths.
Lengths are measured in terms of numbers of characters. 
In both cases, the distributions are centered.
However, the lengths of the tokens in the compressed text are typically larger than those of the original words, as well as more dispersed.

\begin{figure}
\centering
\includegraphics{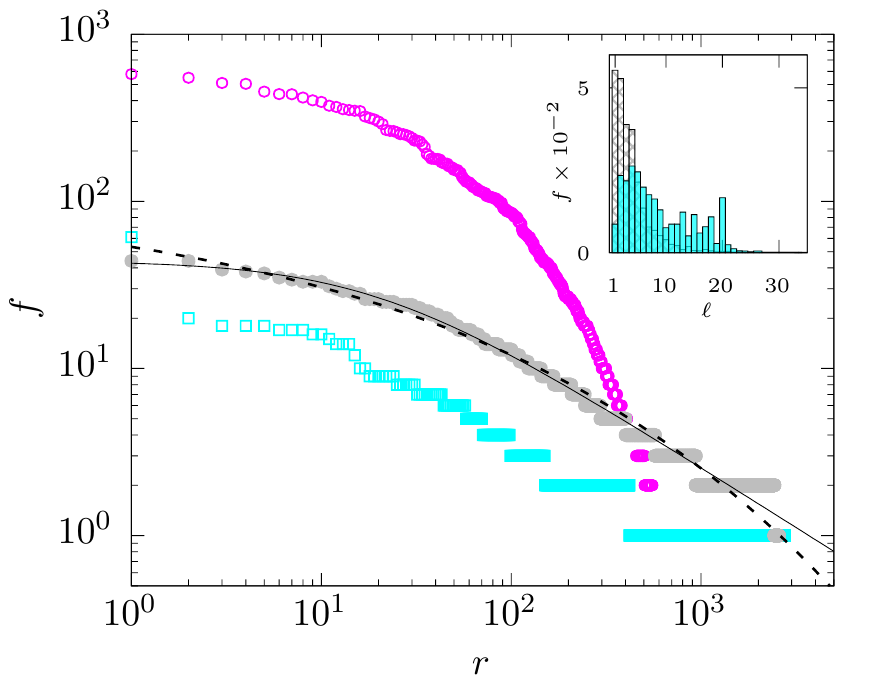}
\caption{
\label{fig:fig6} 
(Color online)
In gray filled circles, the frequency $f$ {\it vs} rank $r$ distribution for the compressed sequence of notes.
For visual comparison, the open magenta circles and cyan squares repeat corresponding distributions from Fig.~\ref{fig:fig1}.
The solid line is the best-fitting ZM function for the distribution of the compressed sequence, with $z=0.726 \pm 0.006$  and  $r_z=14.1 \pm 0.4$. 
The dashed line is the best-fitting SE function with $\xi=0.245 \pm 0.004$ and $r_0=2.9 \pm 0.3$.
Inset: length distributions of g-chords and tokens of the compressed sequence of notes in filled cyan and gray pattern, respectively.
}
\end{figure}

\begin{figure}
\centering
\includegraphics{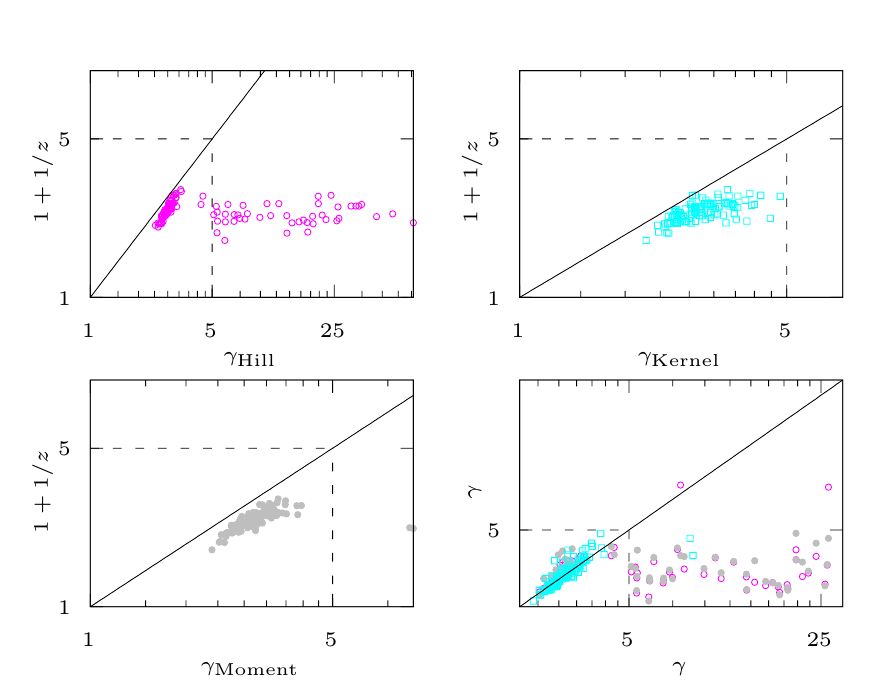}
\caption{
\label{fig:fig7} 
(Color online)
Analogous to Fig.~\ref{fig:fig3}, but for the case of compressed sequences of notes.
A $\%59$ of the compressed sequences pass VT, indicating that compression tends to recover the ZM law.
}
\end{figure}

In Fig.~\ref{fig:fig6}, we repeat the previous study but now using the musical score already analyzed in Fig.~\ref{fig:fig1}.
The curves in magenta open circles and cyan open squares are repeated from Fig.~\ref{fig:fig1} for comparison.
The curve of gray filled circles represents the frequency {\it vs} rank distribution of tokens of the compressed sequence of notes.
One can see that the compressed sequence of notes is a long-tailed distribution, similar to the distribution of g-chords, but with a slightly smaller ZM exponent $z$.
The SE provides a similar fitting quality than the ZM,
hindering the possibility of discerning between the power-law scaling and the exponential form.
Therefore, it is convenient to complement the analysis with the VT, as we previously did in Figs.~\ref{fig:fig3}~and~\ref{fig:fig4} with the uncompressed sequences.
In Fig.~\ref{fig:fig7}, we show the results of VT with the compressed sequences of notes.
Up to $\% 59$ of the cases pass VT, a smaller but similar percentage to the one found for g-chords.
In this sense, g-chords seem to be more appropriate Zipfian units than the compressed notes, but compressed notes are much better than notes alone.

\subsection{
\label{subsec:heaps}
Heaps' law
}

Finally, to complement and validate the analysis of the characteristics of the Zipfian units in music, let us investigate the validity of Heaps' law~\cite{Font-Clos2015PRL,lu2010zipf} in the musical sequences.
Heaps' law---originally discovered studying the growth of vocabulary in literary works---states the scaling relation $V\sim i^{\nu}$ for the number $V$ of different tokens that can be found within the first $i$ elements of a sequence.
The relation between vocabulary growth and Zipf's law is generally nontrivial~\cite{Font-Clos2015PRL}.
In broad terms, Zipf's law cannot occur with a limited vocabulary growth and, under certain but common conditions, Heaps' and Zipf's exponents asymptotically follow the relation $\nu \approx 1/z$~\cite{lu2010zipf}. 

\begin{figure}
\centering
\includegraphics{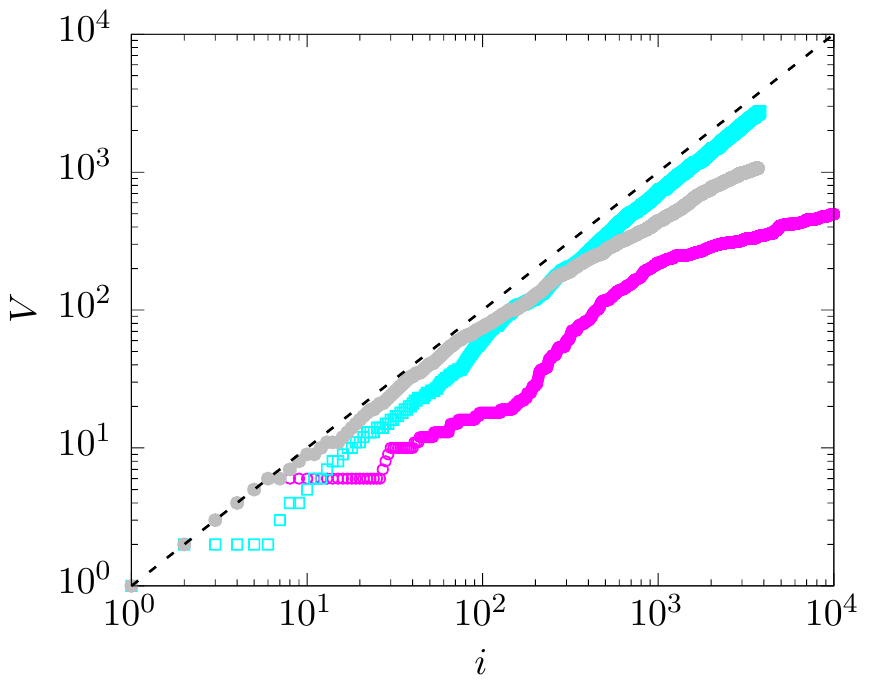}
\caption{
\label{fig:fig8} 
(Color online)
Heaps' curves corresponding to the sequences used to obtain the curves in Fig.~\ref{fig:fig6}.
The dashed line corresponds to a linear scaling $V\sim i$ and is included as a guide to the eye.
}
\end{figure}

In Fig.~\ref{fig:fig8}, curves of $V$ {\it vs} $i$ are plotted in log-log scale for the same sequences analyzed in Fig.~\ref{fig:fig6}.
For g-chords, a linear growth is found, supporting the validity of Heaps law and its consistency with Zipf's law~\cite{lu2010zipf}.
The sequence of notes exhibits a sub-linear growth in the log-log scale, which is the expected behavior since the number of existing notes is bounded. 
The grow of the vocabulary is erratic for the case of notes, partly due to the arbitrary alphabetical sorting.
The result for notes is consistent with the previous analysis showing the failure of Zipf's law for this case.
Finally, for the compressed sequence of notes, the growth of $V$ closely follows Heaps' law, which is consistent with the validity of Zipf's law for the algorithmically obtained Zipfian units.

\section{
\label{sec:discussion}
DISCUSSION
}

The presented statistical analysis indicates that the laws of ZM and Heaps~\cite{Altmann2009,lu2010zipf,Perotti2013} hold for musical scores when g-chords are used as Zipfian units, while non-power-law functional forms provide better fits in the case of notes.
We remark that the presented results for the case of notes are in discrepancy with previously reported findings~\cite{zanette2006zipf}, where the ZM law was observed for a few musical scores of relatively shorter length.
The results here presented, however, involve more than 100 musical scores, rely on the statistical validation of the power-law scaling with the sophisticate VT, include a comparative analysis against literary works and different reference models, and resemble the results reported in~\cite{beltrndelRo2008universality}.
Combined, all these results convey ample support for the statistical validation or rejection of the ZM law in musical scores for different choices of Zipfian units.

The study of sequences of fragmented words---i.e. sequences mimicking a syllabic decomposition of ordinary texts---suggests an interpretation of the presented empirical findings.
The analysis indicates that while g-chords play a similar role to Zipfian units such as words, notes tend to behave like sub-Zipfian units such as word-fragments.
These results are reinforced by those obtained using the nGCA and related reference models.
Specifically, the nGCA can transform sequences of sub-Zipfian units, like notes or word-fragments, into sequences obeying Zipf's and Heaps' laws.
Additionally, it often recovers some of the original Zipfian units.
Furthermore, it frequently produces some sort of super-Zipfian units, reflecting a lurking over-compression tendency.
The over-compression conduces to Zipf's exponents smaller than the originals, agreeing with previous findings where more flatten rank-frequency distributions are observed when n-grams are used instead of words~\cite{ha2002extension,ha2003extension,ha2006reduced,Chierichetti2017}.
In this regard, the observed over-compression tendency can be associated with the existence of repeating n-grams that only occur in particular literary or musical pieces that do not necessarily belong to the general patterns of the underlying written or musical languages.
These results seem related to the evolution~\cite{baixeries2013evolution} and the organization~\cite{bian2016scaling} of language, providing an information-theoretic explanation of the observation of $z < 1$.
We remark, however, that anomalous Zipf's exponents are also observed in other contexts not related to languages~\cite{dasilva2018lotka}.

\section{
\label{sec:conclusions}
CONCLUSIONS
}

This work provides empirical evidence supporting the validity of Zipf's law in musical scores when a proxy for chords is used to define Zipfian units. It also supports the rejection of the use of bare notes because they result in non-Zipfian rank-frequency distributions. 
The Zipf's exponents $z$ we have found for the g-chords are often lesser than one, a result in agreement with the findings by Serra et al.~\cite{serra2012measuring} in the analysis of the harmonic content of western popular music.
These findings are statistically probed using three validation methods in more than 100 musical scores, and complemented with a comparative analysis using literary works.
Combining the empirical statistical analysis with different reference models, we have found that the laws of Zipf and Heaps emerge---at least approximately---from the actual compression of the regularities found in sequences of sub-Zipfian units such as notes or word-fragments.
To the extent of our knowledge, our work is the first one providing evidence that the compression of a sequence of correlated tokens results in the emergence of Zipf's law.

In summary, our work supports the presence of Zipf's law in musical scores, endorsing the theory that Zipf's law is a general trait of human languages.
It also suggests that Zipf's law emerges as a consequence of languages evolving into more efficient or complex forms of communication~\cite{baixeries2013evolution}, although we remark that several alternative explanations have been also proposed~\cite{piantadosi2014zipf}.

\section{
\label{sec:acknowledgments}
Acknowledgments
}
We thank 
M. Montemurro 
and 
N. Almeira 
for their useful comments.
Funding: This  work  was  partially  supported by grants from CONICET (PIP 112 20150 10028), 
FonCyT (PICT-2017-0973),  
SeCyT-UNC (Argentina)  
and  
MinCyT C\'ordoba (PID PGC 2018).




\providecommand{\noopsort}[1]{}\providecommand{\singleletter}[1]{#1}%


\appendix

\section{Constructing sequences of g-chords and notes}
\label{app:A}

Generally speaking, a musical score $M$ comprising multiple instruments can be reduced to a multi-set of time-stamped single notes $s_t$ and chords $c_t$.
Here $t$ denotes ordinary time measured in units of crotchet or similar.
Since a chord $c$ is a set of notes played simultaneously---and by the same instrument---then $M$ can be eventually reduced to a multi-set of time-stamped notes $n_t$, each of which either comes from single note $s_t$ or as the constituent of a chord $c_t$.
A g-chord $g_t$ is the multi-set of all notes $n_t$ occurring at time $t$ in $M$.
Hence, $M$ can be further reduced to a time ordered sequences of $g_{t_1},g_{t_2},...$ of g-chords where $t_i<t_j$ if and only if $i<j$.
Finally, a time respecting sequence $g_1,g_2,...$ can be straightforwardly obtained from the sequence $g_{t_1},g_{t_2},...$ by setting $g_i:=g_{t_i}$.
The time respecting sequences $g_1,g_2,...$ are the ones used in the present work.

A time stamped sequence $x_{t_1},x_{t_2},...$ of arbitrary entities $x$ is said to be time sorted if $t_i < t_j$ whenever $i<j$.
In this context, $x_1,x_2,...$ is called the time respecting sequence of $x_{t_1},x_{t_2},...$~.

Given a time respecting sequence of g-chords $g_1,g_2,...$~and some arbitrary ordering for notes, then a corresponding ordered sequence of notes $n_1,n_2,...$ can be easily defined by the following pairwise relation of order: assuming $n_q\in g_i$ and $n_r\in g_j$ for some g-chords $g_i$ and $g_j$, then $n_q<n_r$ if and only if $i<j$ or the predefined ordering applies when $i=j$.
In the present work, the elected arbitrary and predefined ordering is the alphabetical sorting of notes played simultaneously.

\section{Data gathering and pre-processing}
\label{app:B}

\begin{figure}
\centering
\includegraphics[width=11cm]{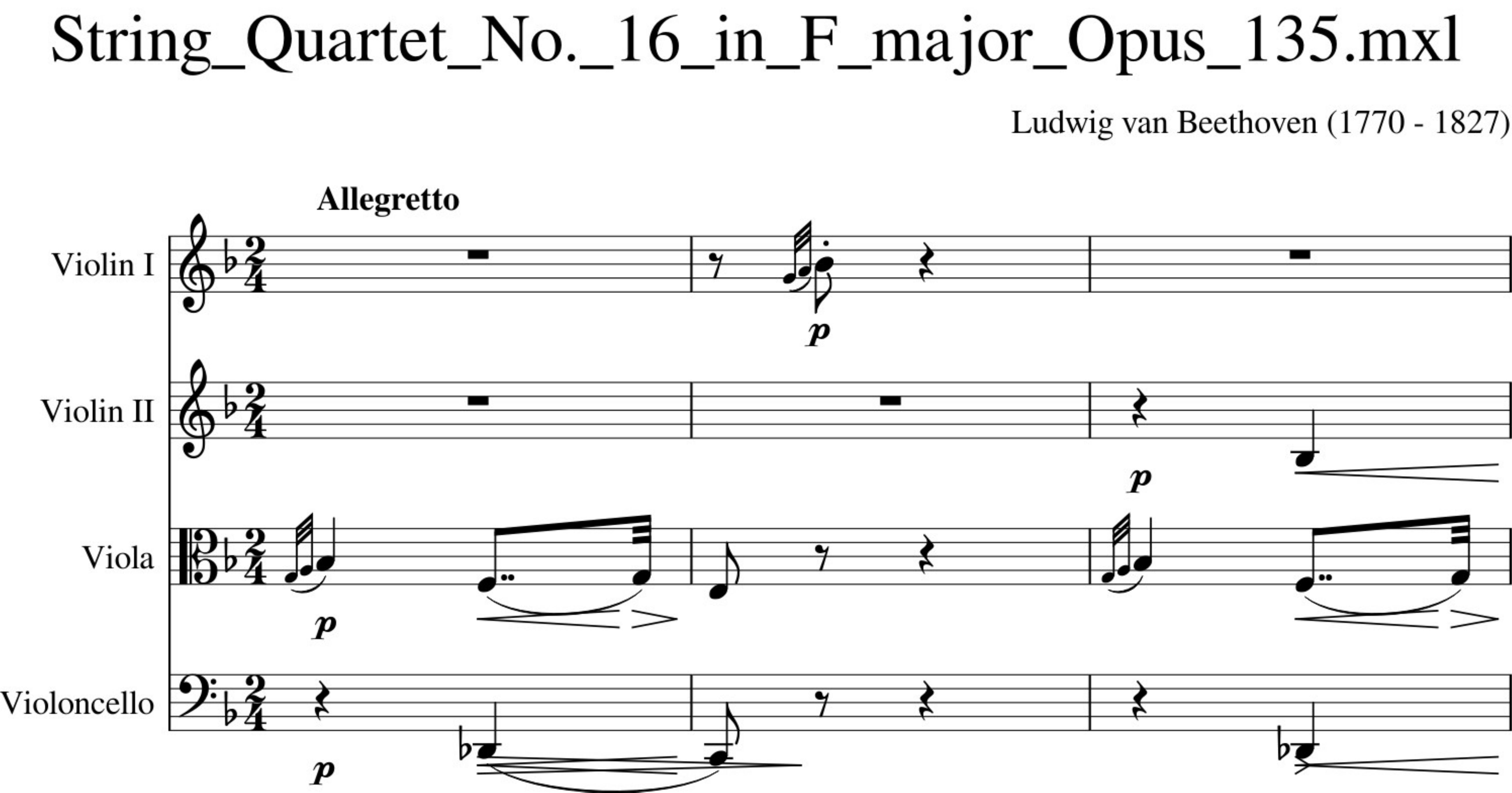}
\cprotect\caption{
\label{fig:fig9} 
The beginning of Beethoven's String Quartet No. 16 in F major, Opus 135, plotted with \verb+Music21 Python+'s package~\cite{music21} and obtained from a corresponding \verb+.mxl+ file format~\cite{mxlformat} that can be downloaded from \verb+MuseScore+~\cite{musescore}.
}
\end{figure}

The musical scores studied in this work were obtained from \verb+MuseScore+~\cite{musescore} as files in \verb+.mxl+ format~\cite{mxlformat}.
We read and pre-processed these files using \verb+Music21 Python+'s package~\cite{music21}.
As an example, by using \verb+Python+'s code, the beginning of Beethoven's String Quartet No. 16 in F major, Opus 135 shown in Fig.~\ref{fig:fig9} can be transformed into the following time-stamped sequences of notes (left) and g-chords (right):
\begin{verbatim}
  time   notes                     time     g-chords

  0.0    '0.0:A3'                  0.0    '0.0:A3&1.0:B-3&0.0:G3'
  0.0    '1.0:B-3'                 1.0    '1.0:D-2&0.875:F3'
  0.0    '0.0:G3'                  1.875  '0.125:G3'
  1.0    '1.0:D-2'                 2.0    '0.5:C2&0.5:E3'
  1.0    '0.875:F3'                2.5    '0.0:A4&0.5:B-4&0.0:G4'
  1.875  '0.125:G3'                4.0    '0.0:A3&1.0:B-3&0.0:G3'
  2.0    '0.5:C2'                  5.0    '1.0:B-3&1.0:D-2&0.875:F3  
  2.0    '0.5:E3'                  ...
  2.5    '0.0:A4'
  2.5    '0.5:B-4'
  2.5    '0.0:G4'
  4.0    '0.0:A3'
  4.0    '1.0:B-3'
  4.0    '0.0:G3'
  5.0    '1.0:B-3'
  5.0    '1.0:D-2'
  5.0    '0.875:F3'    
  ...
\end{verbatim}
Here, notes are represented by two parts, the time duration and the pitch.
For instance, in \verb+1.0:B-3+, \verb+1.0+ represents the time duration and \verb+B-3+ the pitch.
Similarly, g-chords are represented by a sequence of alphabetically sorted notes, separated by the arbitrarily chosen symbol \verb+&+.
Following the common practice in the study of written text, we ignore any punctuation effect that may exist in the musical scores.
All the musical scores studied in the present manuscript were pre-processed in the same manner.
The list of all \verb+.mxl+ files used in our study can be found in a \verb+git+'s repository~\cite{perotti2019zipf_music_gitcode} where,
additionally, \verb+Python+'s code with pre-processing examples and runs of the nGCA can be also found.

We do not offer within the repository the \verb+.mxl+ files nor the corresponding sequences because of copyright issues.
On the other hand, they can be downloaded by the reader or we can share them upon request.

\end{document}